\documentclass[aps,prl,twocolumn,groupedaddress]{revtex4}
\usepackage[pdftex]{hyperref}
\usepackage{graphicx}
\usepackage{subfigure}
\usepackage{verbatim}

\begin{document}

\title{Serially-regulated biological networks fully realize a constrained set of functions}

\author{Andrew Mugler} \email{ajm2121@columbia.edu}
\affiliation{Department of Physics, Columbia University, New York, NY
  10027} \author{Etay Ziv} \affiliation{College of Physicians and
  Surgeons, Columbia University, New York, NY 10027} \author{Ilya
  Nemenman} \affiliation{Computer, Computational and Statistical
  Sciences Division, Center for Nonlinear Studies, Los Alamos National
  Laboratory, Los Alamos, NM 87545} \author{Chris H. Wiggins}
\affiliation{Department of Applied Physics and Applied Mathematics,
  Center for Computational Biology and Bioinformatics, Columbia
  University, New York, NY 10027}

\date{\today}

\begin{abstract}
  We show that biological networks with serial regulation (each node
  regulated by at most one other node) are constrained to {\it direct
    functionality}, in which the sign of the effect of an
  environmental input on a target species depends only on the direct
  path from the input to the target, even when there is a feedback
  loop allowing for multiple interaction pathways. Using a stochastic
  model for a set of small transcriptional regulatory networks that
  have been studied experimentally \cite{Guet:2002p38}, we further
  find that all networks can achieve all functions permitted by this
  constraint under reasonable settings of biochemical parameters.
  This underscores the functional versatility of the networks.
\end{abstract}

\maketitle

A driving question in systems biology in recent years has been the
extent to which the topology of a biological network determines or
constrains its function. Early works have suggested that the function
follows the topology \cite{ShenOrr:2002p101, Mangan:2003p41,
  Guet:2002p38, Kollmann:2005p103}, and this continues as a prevailing
view even though later analyses (at least in a small corner of
biology) have questioned the paradigm \cite{Wall:2005p20,
  Ziv:2007p88}. It remains unknown if a small biochemical or
regulatory network can perform multiple functions, and whether the
function set is limited by the network's topological structure. To
this extent, in this paper, we develop a mathematical description of
the functionality of a certain type of biological network, and show
that the answer to both questions is ``yes'': the networks can perform
many, but not all possible functions, and the set of attainable
functions is constrained by the topology. We illustrate these results
in the context of an experimentally realized system
\cite{Guet:2002p38}.

Following \cite{Guet:2002p38} and our earlier work \cite{Ziv:2007p88},
we focus on the steady-state functionality of transcriptional
regulatory networks. In this case, the input is the ``chemical
environment,'' that is a binary vector of presence/absence of small
molecules that affect the regulation abilities of the transcription
factors; and the output is the steady-state expression of a particular
gene, hereafter called the reporter. Different functions of the
network correspond then to different ways to map the small molecule
concentrations into the reporter expression.

In our setup, the effect of introducing a small molecule S$_j$
specific to a transcription factor X$_j$ is to modify the affinity of
X$_j$ to its binding site.  Equivalently one can think of S$_j$ as
modulating or renormalizing the transcription factor concentration
$X_j$ by some factor $s_j$, making the effective concentration $\chi_j
= \chi_j(X_j, s_j)$.  A simple example of such a modulation function is
\begin{equation}
\label{eq:chi0}
\chi_j(X_j, s_j) \equiv X_j/s_j,
\end{equation}
in which the presence of the small molecule reduces the effective concentration of transcription factor by the factor $s_j$.

The function of the circuit will depend on how
the steady-state expression $G^*$ of the reporter gene G changes as
the modulation factor $s_j$ is varied from some ``off'' value $s_j^-$ to some
``on'' value $s_j^+$:
\begin{equation}
  \label{eq:deltaG}
  \frac{\Delta G^*}{\Delta s_j} = \frac{G^*(s_j^+) - G^*(s_j^-)}{\Delta s_j}
  = \frac{1}{\Delta s_j}\int_{s_j^-}^{s_j^+} \frac{dG^*}{ds_j} \, ds_j,
\end{equation}
where $\Delta s_j = s_j^+ - s_j^-$.  For example, if $\chi_j = X_j/s_j$, then $s_j^- = 1$, indicating that the small molecule is absent, and $s_j^+ > 1$ is the factor by which effective concentration is reduced when the small molecule is present.

If the sign of $dG^*/ds_j$ does
not change for $s_j\in [s_j^-, s_j^+]$, then the sign of $\Delta G^*$
is fixed.  For networks with only serial regulation, i.e.\ each gene
is regulated by at most one other gene, we will show that the sign of
$dG^*/ds_j$ is unique and in accord with the direct path from S$_j$ to
G, a property we term {\it direct functionality}. This constrains the
possible responses and hence the functionality of serial networks.
Importantly, we will then show that all admissible functions indeed
can be attained by all the networks we studied operating at different
parameter values. While throughout this work we focus on the setup
pioneered experimentally by Guet et al.\ \cite{Guet:2002p38}, we also
show that the constraint to direct functionality holds for any network
with serial regulation.

\section{Direct functionality in small networks}
\label{sec:direct}

\begin{figure}
\centering
\includegraphics [scale=.4] {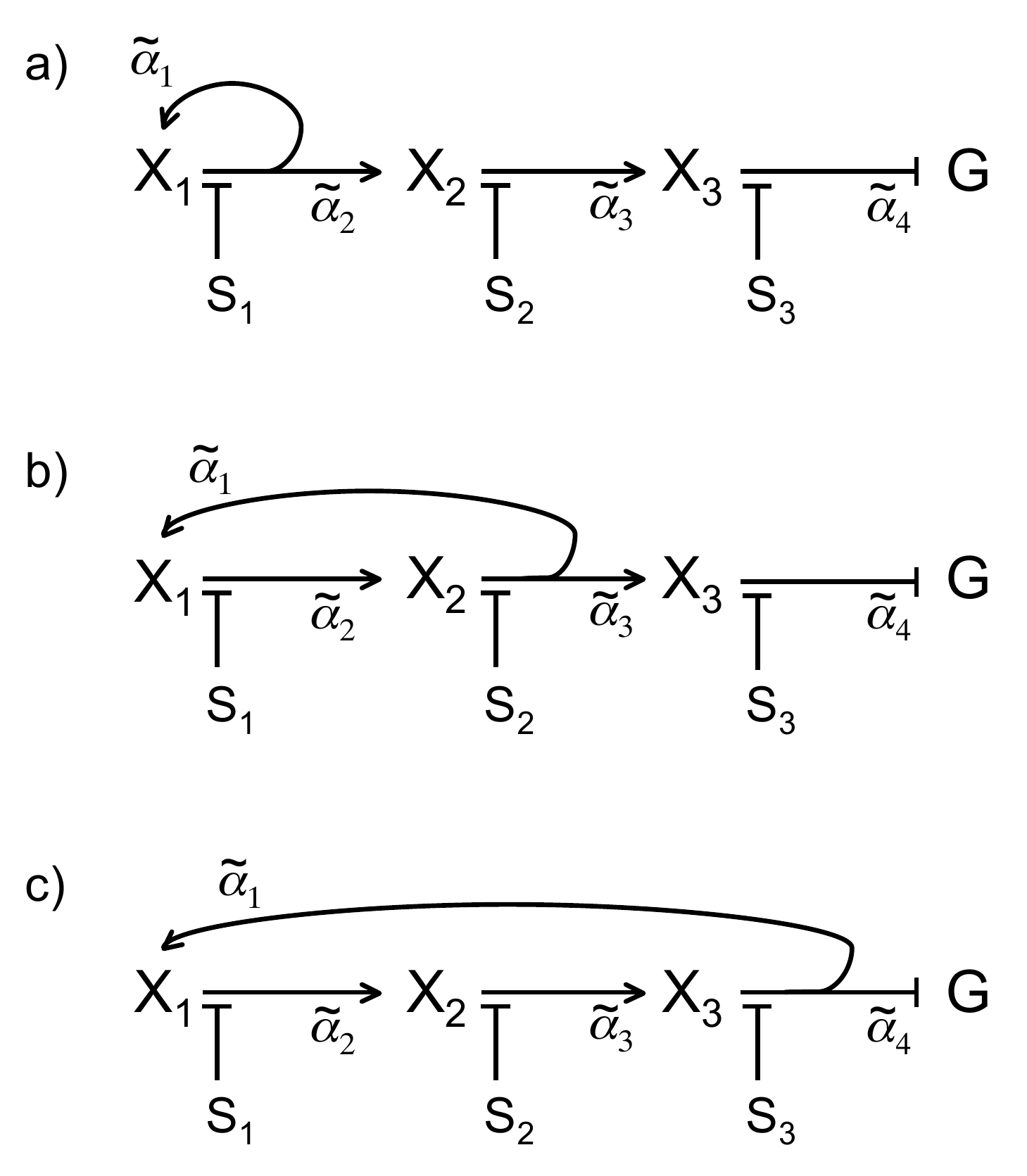}
\caption{Four-gene networks (3 transcription factors X$_i$ plus 1
  reporter gene G) in which each gene is regulated by one other gene,
  as studied in \cite{Guet:2002p38, Ziv:2007p88}.  Transcription
  factor efficacies are influenced by small molecules S$_i$.
  Regulation functions $\tilde{\alpha}_i$ are assigned to the edges.
  The three edges $\tilde{\alpha}_1$, $\tilde{\alpha}_2$, and
  $\tilde{\alpha}_3$ can be up-regulating or down-regulating, giving
  $3 \times 2^3 = 24$ possibilities; the reporter gene is repressed in
  all cases.}
\label{fig:topos}
\end{figure}

As in Guet et al.\ \cite{Guet:2002p38}, we consider networks with $N$
= 4 genes (three transcription factors plus a reporter G), in which
each gene is regulated by exactly one other gene.  This admits three
topologies and a total of 24 networks, as described in Fig.\
\ref{fig:topos}.  All three topologies consist of a cycle and a
cascade that begins in the cycle and ends at the reporter gene G.
Once outside the cycle, there is only one path to G, so it suffices to
study a topology consisting of an $n$-gene cycle with a gene G
immediately outside (Fig.\ \ref{fig:topos}c is an example with $n =
3$), and extensions to topologies where the cycle is connected to the
reporter by a linear cascade are trivial.

In this section, we will perform the steady-state analysis of such
single-cycle networks to lay the groundwork for understanding the effect
of topology on allowed functionality.

The process of protein expression has been modeled with remarkable
success by combining transcription and translation into one step and
directly coupling genes by a deterministic dynamics
\cite{Elowitz:2000p37, Gardner:2000p77, Hasty:2001p84}.  Accordingly
we model mean expressions $\bar{X}_i$ (we later distinguish between entire probability distributions $P(X_i)$ and the means of these distributions $\bar{X}_i$; cf.\ Appendix) with the system of ordinary
differential equations
\begin{eqnarray}
\label{eq:ode1}
\frac{d\bar{X}_1}{dt} &=& \tilde{\alpha}_1(\bar{\chi}_n) - r_1 \bar{X}_1,\\
\label{eq:odei}
\frac{d\bar{X}_i}{dt} &=& \tilde{\alpha}_i(\bar{\chi}_{i-1}) - r_i \bar{X}_i \quad (2 \leq i \leq n),\\
\label{eq:odeG}
\frac{d\bar{G}}{dt} &=& \tilde{\alpha}_{n+1}(\bar{\chi}_n) - r_{n+1} \bar{G},
\end{eqnarray}
where the $\tilde{\alpha}_i$ are creation rates for the species X$_i$ (and $X_{n+1} \equiv G$),
each monotonically regulated by the effective concentration
$\bar{\chi}_{\pi_i}$ of its parent $\pi_i$, and the $r_i$ are the
decay rates. Note that we have set
\begin{eqnarray}
\pi_1 &=& n, \\
\pi_i &=& i-1 \quad (2 \leq i \leq n+1)
\end{eqnarray}
to create the $n$-gene cycle with one gene immediately outside.  The regulation functions $\tilde{\alpha}_i$ will be up- or down-regulating according to the network topology.  A common example is the familiar Hill functions,
\begin{eqnarray}
\label{eq:hill1_0}
\tilde{\alpha}(\bar{\chi}) &=& a_0 + a\frac{\bar{\chi}^h}{K^h + \bar{\chi}^h}
\quad \textrm{(up-regulating)},\\
\label{eq:hill2_0}
\tilde{\alpha}(\bar{\chi}) &=& a_0 + a\frac{K^h}{K^h + \bar{\chi}^h}
\quad \textrm{(down-regulating)},
\end{eqnarray}
with basal and maximal expression levels $a_0$ and $a_0 + a$
respectively, Michaelis-Menten constants $K$, and cooperativities $h$.  Although we use the functional forms in Eqns.\ (\ref{eq:hill1_0}-\ref{eq:hill2_0}), as well as the functional form for the modulation function in Eqn.\ (\ref{eq:chi0}), for our numerical experiment (cf.\ Numerical Results), the analytic result derived in this section will be valid for any monotonic functions $\tilde{\alpha}(\bar{\chi})$ and any function $\bar{\chi}(\bar{X}, s)$.

Fixed points of the dynamical system in Eqns.\ (\ref{eq:ode1}-\ref{eq:odeG}) satisfy
\begin{eqnarray}
\label{eq:fp1}
\bar{X}_1^* &=& \alpha_1(\bar{\chi}_n^*),\\
\bar{X}_i^* &=& \alpha_i(\bar{\chi}_{i-1}^*) \quad (2 \leq i \leq n),\\
\label{eq:fp3}
\bar{G}^* &=& \alpha_{n+1}(\bar{\chi}_n^*),
\end{eqnarray}
where we define
\begin{equation}
\alpha_i \equiv \tilde{\alpha}_i/r_i.
\end{equation}

We may now, as in
\cite{Kholodenko:1997p55, Kholodenko:2002p43}, use the chain rule to
calculate the derivative of $\bar{G}^*$ with respect to a particular
input factor $s_j$.  For illustration, we will do so first for the concrete example in Fig.\ \ref{fig:topos}c, in which $n=3$.  Let us consider the derivative of $\bar{G}^*$ with respect to $s_1$:
\begin{equation}
\label{eq:3chain1}
\frac{d\bar{G}^*}{ds_1} = \frac{\partial\alpha_4}{\partial \bar{X}_3} \frac{\partial\alpha_3}{\partial \bar{X}_2} \left[\frac{\partial\alpha_2}{\partial s_1}+\frac{\partial\alpha_2}{\partial \bar{X}_1}\frac{d\bar{X}_1^*}{d s_1}\right],
\end{equation}
where all derivatives are evaluated at the fixed point, and it is understood that $\alpha_i$ depends on either $\bar{X}_{\pi_i}$ or $s_{\pi_i}$ through $\bar{\chi}_{\pi_i}$, that is, that
\begin{equation}
\label{eq:chiXs}
\frac{\partial\alpha_i}{\partial \bar{X}_{\pi_i}} = \frac{\partial\alpha_i}{\partial \bar{\chi}_{\pi_i}}
	\frac{\partial\bar{\chi}_{\pi_i}}{\partial \bar{X}_{\pi_i}} \quad {\rm and} \quad
\frac{\partial\alpha_i}{\partial s_{\pi_i}} = \frac{\partial\alpha_i}{\partial \bar{\chi}_{\pi_i}}
	\frac{\partial\bar{\chi}_{\pi_i}}{\partial s_{\pi_i}}.
\end{equation}
If we introduce the notation
\begin{eqnarray}
\alpha_i' &\equiv& \partial\alpha_i/\partial \bar{X}_{\pi_i}, \\
\dot{\alpha_i} &\equiv& \partial \alpha_i/\partial s_{\pi_i},
\end{eqnarray}
then Eqn.\ (\ref{eq:3chain1}) becomes 
\begin{equation}
\label{eq:3chain2}
\frac{d\bar{G}^*}{ds_1} = \alpha'_4 \alpha'_3 \left[ \dot{\alpha}_2 + \alpha'_2 \frac{d\bar{X}_1^*}{ds_1} \right].
\end{equation}
The first term reflects the direct chain to G from S$_1$, and the second term incorporates further contributions around the cycle and will need to be evaluated self-consistently.

For a cycle of arbitrary length $n$ and for an arbitrary input factor $s_j$ ($1 \le j \le n$), Eqn.\ (\ref{eq:3chain2}) generalizes to
\begin{equation}
\label{eq:dGds}
\frac{d\bar{G}^*}{ds_j} = \left[ \dot{\alpha}_{j+1} + \alpha'_{j+1} \frac{d\bar{X}_j^*}{ds_j} \right]
	\prod_{k=j+2}^{n+1} \alpha'_k,
\end{equation}
where we use the convention that
\begin{equation}
\prod_{k=a}^b [\cdot] = 1 \quad {\rm if} \quad a>b.
\end{equation}
We may also use the chain rule for $d\bar{X}_j^*/ds_j$,
\begin{eqnarray}
\frac{d\bar{X}_j^*}{ds_j} &=&
	\left[ \dot{\alpha}_{(j \, {\rm mod} \, n)+1}
	+ \alpha'_{(j \, {\rm mod} \, n)+1} \frac{d\bar{X}_j^*}{ds_j} \right] \nonumber \\
	&& \times \frac{\prod_{k=1}^n \alpha'_k}{\alpha'_{(j \, {\rm mod} \, n)+1}}
\end{eqnarray}
and now we may solve for $d\bar{X}_j^*/ds_j$ self consistently:
\begin{equation}
\label{eq:dXjdsj}
\frac{d\bar{X}_j^*}{ds_j} =
	\frac{\dot{\alpha}_{(j \, {\rm mod} \, n)+1}
	}{\alpha'_{(j \, {\rm mod} \, n)+1}} \frac{\prod_{k=1}^n \alpha'_k}{
	1 - \prod_{l=1}^n \alpha'_l}.
\end{equation}
For the special case of $j=n$, where $(j \, {\rm mod} \, n)+1 = 1$, substituting Eqn.\ (\ref{eq:dXjdsj})
into Eqn.\ (\ref{eq:dGds}) obtains
\begin{eqnarray}
\frac{d\bar{G}^*}{ds_n} &=&
	 \left[ \frac{1}{1 - \prod_{l=1}^n \alpha'_l} \right] \nonumber \\
	&& \times \left[ \dot{\alpha}_{n+1} + ( \dot{\alpha}_1 \alpha'_{n+1} - 
	\dot{\alpha}_{n+1} \alpha'_1 ) \prod_{k=2}^n \alpha'_k \right]\\
\label{eq:dGdsn}
	&=& \left[ \frac{1}{1 - \prod_{l=1}^n \alpha'_l} \right] \dot{\alpha}_{n+1},
\end{eqnarray}
where the second step follows from
\begin{eqnarray}
\dot{\alpha}_1 \alpha'_{n+1}
	&=& \left( \frac{d\alpha_1}{d\bar{\chi}_n}\frac{\partial \bar{\chi}_n}{\partial s_n} \right)
	\left( \frac{d\alpha_{n+1}}{d\bar{\chi}_n}\frac{\partial \bar{\chi}_n}{\partial \bar{X}_n} \right)\\
	&=& \left( \frac{d\alpha_1}{d\bar{\chi}_n}\frac{\partial \bar{\chi}_n}{\partial \bar{X}_n} \right)
	\left( \frac{d\alpha_{n+1}}{d\bar{\chi}_n}\frac{\partial \bar{\chi}_n}{\partial s_n} \right)\\
	&=& \alpha'_1 \dot{\alpha}_{n+1},
\end{eqnarray}
in which the first step recalls Eqn.\ (\ref{eq:chiXs}).  For $1 \leq j \leq n-1$, where $(j \, {\rm mod} \, n)+1 = j+1$, substituting Eqn.\ (\ref{eq:dXjdsj}) into
Eqn.\ (\ref{eq:dGds}) obtains
\begin{equation}
\label{eq:dGds_1}
\frac{d\bar{G}^*}{ds_j} = \left[ \frac{1}{1 - \prod_{l=1}^n \alpha'_l} \right]
	\dot{\alpha}_{j+1} \prod_{k=j+2}^{n+1} \alpha'_k,
\end{equation}
which, upon inspection of Eqn.\ (\ref{eq:dGdsn}), is valid for $j=n$
as well.

Stability of the fixed point $\bar{X}^*_j$ requires that the Jacobian
of Eqns.\ (\ref{eq:ode1}-\ref{eq:odei}),
\begin{equation}
\label{eq:jacobian}
J = \left[
	\begin{array}{cccccc}
	-r_1 &&&&& \tilde{\alpha}'_1 \\
	\tilde{\alpha}'_2 & -r_2 &&&& \\
	& \tilde{\alpha}'_3 & -r_3 &&& \\
	&& \ddots & \ddots && \\
	&&& \tilde{\alpha}'_{n-1} & -r_{n-1} & \\
	&&&& \tilde{\alpha}'_n & -r_n \\
	\end{array}
	\right],
\end{equation}
be negative definite or, since the determinant is the product of the
eigenvalues, that
\begin{eqnarray}
0 &<& (-1)^n\det(J)\\
	&&= \prod_{k=1}^n r_k - \prod_{l=1}^{n} \tilde{\alpha}'_l\\
\label{eq:stability}
	&&= \prod_{k=1}^n r_k \left( 1 - \prod_{l=1}^{n} \alpha'_l \right).
\end{eqnarray}
Since the decay rates $r_k$ are positive, Eqn.\ (\ref{eq:stability})
says that the term inside the brackets in Eqn.\ (\ref{eq:dGds_1}) is
positive for stable fixed points.

For the networks in Fig.\ \ref{fig:topos}, where in the 1- and
2-cycles the reporter is attached by means of intermediates, the
analog of Eqn.\ (\ref{eq:dGds_1}) is calculated similarly to be
\begin{equation}
\label{eq:dGds24}
\frac{d\bar{G}^*}{ds_j} = \left[ \frac{1}{1 - \theta(n-j) \prod_{l=1}^n \alpha'_l} \right]
	\dot{\alpha}_{j+1} \prod_{k=j+2}^{N} \alpha'_k,
\end{equation}
where $N=4$ is the number of genes, $1 \le j \le N-1$ for each of the 3 possible small molecule inputs, and $n$ is the length of the cycle ($1 \le n \le N-1$).  Here $\theta$ is the Heaviside function, for which we use the convention $\theta(0) = 1$.  Its presence reduces the bracketed term to 1 when the input S$_j$ is outside the cycle, leaving only the contribution corresponding to the cascade from S$_j$ to G, as must be the case.

In Eqn.\ (\ref{eq:dGds24}), the term outside
the brackets represents the direct (i.e., the shortest) path from
S$_j$ to G and fixes the sign of $d\bar{G}^*/ds_j$ (since the term
inside the brackets is positive at a stable fixed point).  If the
creation rates are monotonic (which is the usual model for
transcriptional regulation, but may be violated in protein signaling
due to competitive inhibition and other effects), this sign is unique
and fixes the sign of $\Delta \bar{G}^*/\Delta s_j$ via Eqn.\
(\ref{eq:deltaG}).  Importantly, this says that the feedback in each
of the topologies in Fig. \ref{fig:topos} is irrelevant in determining
the sign of $\Delta \bar{G}^*/\Delta s_j$ for a steady-state analysis.
As an example, for the network in Fig.\ \ref{fig:func}a (inset), $\bar{G}^*$ changes
with increasing $s_1$ according to $\dot{\alpha}_2\alpha'_3\alpha'_4$,
which, since S$_1$ inhibits the activation, is negative $\times$
positive $\times$ negative $=$ positive, just as one would expect if
the feedback was ignored.

\subsection{Direct functionality corresponds to specific orderings of
  output states}

Consider the case in which there are only two small molecule inputs,
S$_1$ and S$_2$, as in Fig.\ \ref{fig:func}a (inset).  Since each
input can be absent or present, $S_1, S_2 \in \{-, +\}$, there are
four chemical input states $c = S_1S_2 \in \{--, -+, +-, ++\}$.
Direct functionality admits only two orderings of the four output
states $\bar{G}^*_c$, and hence the functionality of the network is
severely limited by its serial topology.  To see this, note that for
Fig.\ \ref{fig:func}a (inset) we have
\begin{eqnarray}
\Delta \bar{G}^*/\Delta s_1 \geq 0 &\Rightarrow& \bar{G}^*_{+-} \geq \bar{G}^*_{--}
	\textrm{ and} \nonumber \\
	&& \bar{G}^*_{++} \geq \bar{G}^*_{-+};\\
\Delta \bar{G}^*/\Delta s_2 \geq 0 &\Rightarrow& \bar{G}^*_{-+} \geq \bar{G}^*_{--}
	\textrm{ and} \nonumber \\
	&& \bar{G}^*_{++} \geq \bar{G}^*_{+-}.
\end{eqnarray}
These conditions permit only the following output orderings,
irrespective of biochemical parameters:
\begin{eqnarray}
\bar{G}^*_{--} &\leq& \bar{G}^*_{-+} \leq \bar{G}^*_{+-} \leq \bar{G}^*_{++}
	\textrm{ or} \nonumber \\
\label{eq:2ords}
\bar{G}^*_{--} &\leq& \bar{G}^*_{+-} \leq \bar{G}^*_{-+} \leq \bar{G}^*_{++}.
\end{eqnarray}
These two orderings nevertheless allow the realization of a
significant subset of all possible logical functions that one can
build with two binary inputs, depending on the distinguishability of
the four output states, as described in the next section.  Quantifying
the distinguishability demands careful treatment of the
noise with a stochastic equivalent of our deterministic dynamical
system, as described in the Appendix.

\section{Numerical results}
\label{sec:results}

We numerically solved the system in Eqns.\
(\ref{eq:ode1}-\ref{eq:odeG}) [with stochastic effects given by Eqn.\
(\ref{eq:lyap})] with many parameter settings for all 24 networks
represented in Fig.\ \ref{fig:topos}.  In addition to verifying the
restriction to direct functionality, we find that all networks can
achieve all possible direct functions, suggesting that the networks
are still quite versatile within the functional constraint.

\begin{figure}
\centering
\subfigure[]{\includegraphics [scale=.29] {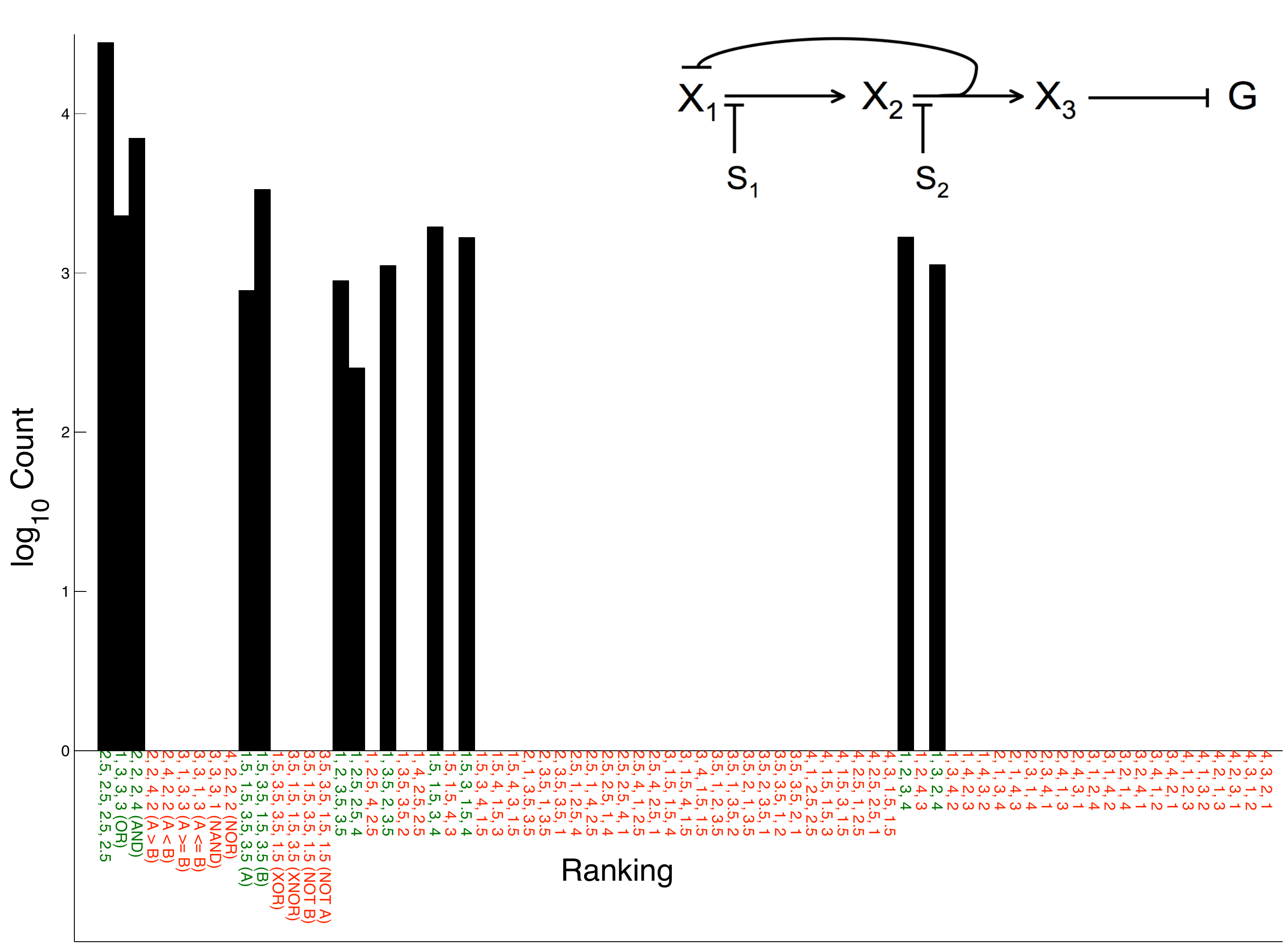} }
\subfigure[]{\includegraphics [scale=.28] {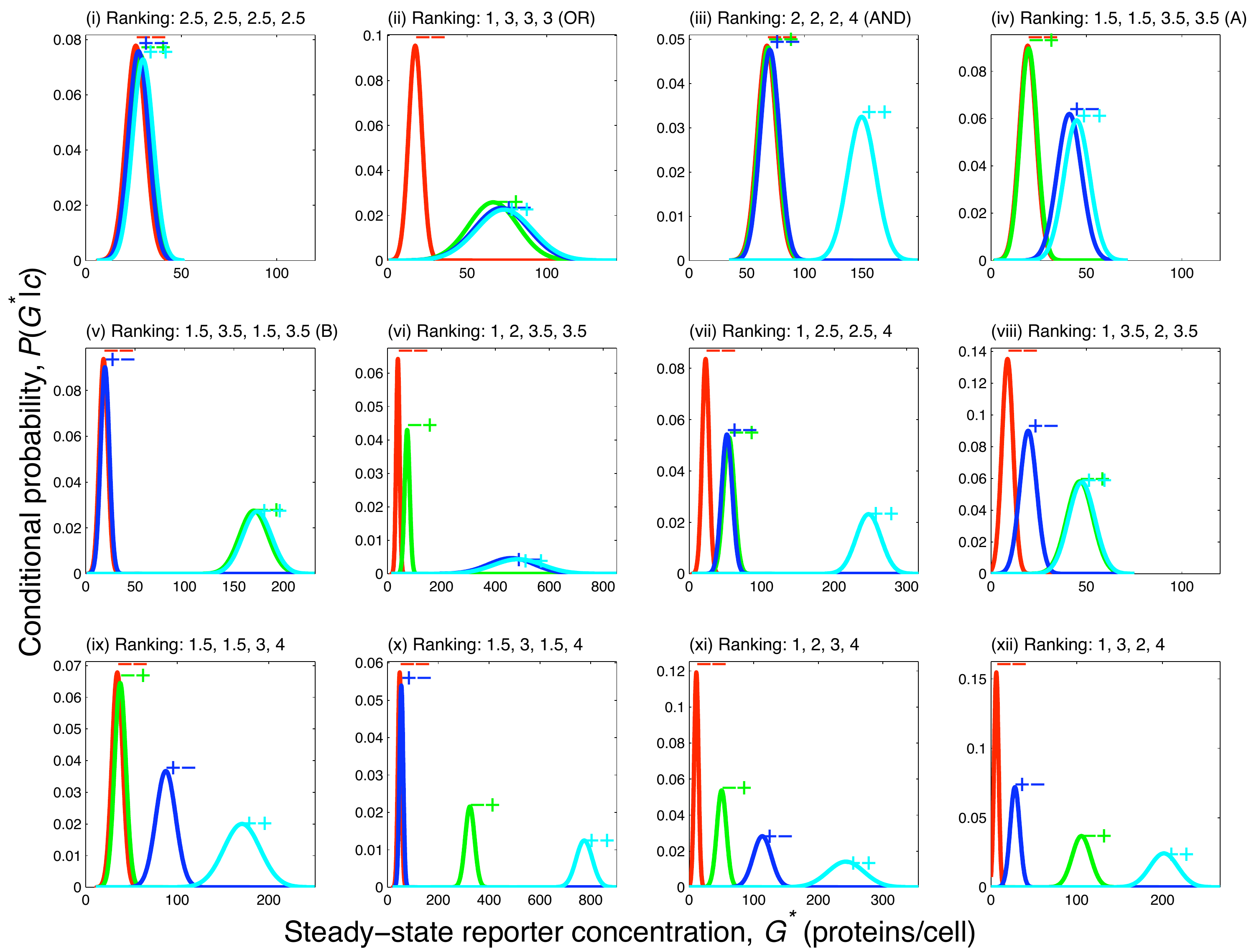} }
\caption{Direct functionality in a representative network with serial
  regulation (network shown in the inset in (a)).  {\bf (a)} Histogram
  of logical functions, as defined by the ranking of the output
  distributions $P(G^*|c)$ (cf.\ Numerical Results).  Binary logic
  names are included after rankings when applicable, with `A' and `B'
  corresponding to inputs S$_1$ and S$_2$ respectively.  Direct
  functions are labeled in green, indirect functions in red.  Note
  that only direct functions are observed, and that all direct
  functions can be attained by the network.  {\bf (b)} An example of
  each direct function.  Two distributions are considered
  indistinguishable in rank when their means are separated by less
  than the smaller of their standard deviations.  For example, in
  (ix), the distributions for the first and second states ($--$ and
  $-+$ respectively) are indistinguishable, so they are tied in rank
  at 1.5 (the average of ranks 1 and 2).  Note that all functions
  satisfy the ordering constraints in Eqn.\ (\ref{eq:2ords}).}
\label{fig:func}
\end{figure}

For all networks, we consider the case of two small
molecule inputs S$_1$ and S$_2$, as in the experimental setup of Guet et al.\ \cite{Guet:2002p38}, and as shown for an example network in Fig.\ \ref{fig:func}a (inset). We take $s_j$ to be a multiplicative factor by which the transcription
factor concentration $\bar{X}_j$ is effectively scaled, i.e.\
\begin{equation}
\label{eq:chi}
\bar{\chi}_j(\bar{X}_j, s_j) \equiv \bar{X}_j/s_j.
\end{equation}
Then $s_j^- \equiv 1$ for the ``off'' settings, and the $s_j^+ > 1$
are free parameters for the ``on'' settings.

We model the regulation
using the familiar Hill form (which is monotonic and thus satisfies the
direct functionality conditions)
\begin{eqnarray}
\label{eq:hill1}
\tilde{\alpha}(\bar{\chi}) &=& a_0 + a\frac{\bar{\chi}^h}{K^h + \bar{\chi}^h}
\quad \textrm{(up-regulating)},\\
\label{eq:hill2}
\tilde{\alpha}(\bar{\chi}) &=& a_0 + a\frac{K^h}{K^h + \bar{\chi}^h}
\quad \textrm{(down-regulating)},
\end{eqnarray}
with basal and maximal expression levels $a_0$ and $a_0 + a$
respectively, Michaelis-Menten constants $K$, and cooperativities $h$.
For the 4-gene networks in Fig.\ \ref{fig:topos}, with only two small
molecule inputs S$_1$ and S$_2$, this gives 22 parameters in total
(cf.\ Table \ref{tab:params}).

\begin{table}
\centering
\begin{tabular}{|l|l|}
\hline
Parameters & Range \\ \hline\hline
decay rates, $r_i$ & $10^{-4} - 10^{-3}$ \\ \hline
Michaelis-Menten constants, $K_i$ & $10^0 - 10^3$ \\ \hline
basal expression levels, $a_{0, i}$ & $10^{-3} - 10^{-2}$ \\ \hline
expression level ranges, $a_i$ & $10^0 - 10^2$ \\ \hline
cooperativities, $h_i$ & $10^0 - 10^1$ \\ \hline
``on'' input factors, $s_j^+$ & $10^2 - 10^3$ \\ \hline
\end{tabular}
\caption{Parameters and ranges from which each is randomly drawn, with $1 \leq i \leq 4$ for the four genes, and $1 \leq j \leq 2$ for the two small molecule inputs.  Ranges are representative of typical cell conditions \cite{Elowitz:2000p37, weiss}.}
\label{tab:params}
\end{table}

For a given parameter set, we numerically solve Eqns.\
(\ref{eq:ode1}-\ref{eq:odeG}) (using Matlab's \texttt{ode15s}) for
each input state $c \in \{--, -+, +-, ++\}$ to find mean steady-state
concentrations $\bar{G}^*_c$.  We then solve Eqn.\ (\ref{eq:lyap}) to
find fluctuations around these means, giving probability distributions
$P(G^*|c)$ (cf.\ Appendix).  The function is defined by the ranking of
the conditional distributions $P(G^*|c)$.  That is, if two
distributions are distinguishable, then the one with the larger mean
is ranked higher.  We consider two distributions to be
indistinguishable when their means are separated by less than the
smaller of their standard deviations (alternative definitions do not
change our results qualitatively), in which case they both take on the
average of their two ranks.  When there are only two distinguishable
output states, this rank-based classification reduces to that defining
the familiar binary logical functions {\tt AND}, {\tt OR}, {\tt XOR}
etc.\ (see, for example, Fig.\ \ref{fig:func}b, (ii-v)).  More
generally, for one, two, three, and four distinguishable responses,
there are 75 total rankings (as listed on the horizontal axis of
Fig. \ref{fig:func}a).  However, only 12 of these satisfy the ordering
constraints for each network analogous to those in Eqn.\
(\ref{eq:2ords}) and therefore correspond to direct functions (for the
newtork in Fig.\ \ref{fig:func}a these 12 are shown in green on the
horizontal axis).

We ran 50,000 trials for each of the 24 networks, in which the
parameters were randomly selected (using a distribution uniform in
log-space) from the ranges in Table \ref{tab:params}. We found the
steady-state reporter expression distributions $P(G^*|c)$ and
classified the responses by ranking.  All 24 networks displayed only
direct functions.  However, every network was able to achieve all 12
of its direct functions with parameters selected via Table
\ref{tab:params}, meaning that the networks fully realized all the
functionality allowed by the constraint.  This suggests that the networks studied
are both constrained and versatile, and that a cell may still use a
serial network to perform multiple logical functions by varying
biochemical parameters, despite the restriction to direct
functionality.  Fig.\ \ref{fig:func} shows a histogram of functions
and an example of each type of direct function for a representative
network.

We note that Guet et al.\ experimentally observed both direct and
indirect functions \cite{Guet:2002p38}.  However, they explicitly call
the indirect functions into question, citing several possible
unanticipated effects including RNA polymerase read-through.  We have
not incorporated such effects into the current model.

\section{Multiple fixed points}

For the 12 networks in which the overall sign of the feedback cycle is
positive, there are parameter settings that support multiple stable
fixed points.  In this section we evaluate the extent to which the
presence of multiple fixed points affects the constraint to direct
functionality, and we find that violation of the constraint is
possible but unlikely.

While the function of a network has been defined in terms of
$P(G^*|c)$, the linear noise approximation (cf.\ Appendix) only gives
us access to $P(G^*|c, \bar{\bf X}^*_m)$, the distribution expanded
around a particular fixed point $\bar{\bf X}^*_m$.  The two are
related by a weighted sum,
\begin{equation}
P(G^*|c) = \sum_m \pi_m P(G^*|c, \bar{\bf X}^*_m),
\end{equation}
where the probabilities $\pi_m$ of being near the $m$th fixed point
will depend on the basins of attraction and curvatures near the fixed
points.  Numerical solution for $P(G^*|c)$ directly is possible in
principle, although computationally difficult. Whether the statistical
steady state distribution is calculated numerically or is approximated
as in this manuscript, if we continue to define the function of the
network by the ranking of the means of the $P(G^*|c)$, we
have
\begin{eqnarray}
\frac{d\bar{G}^*}{ds_j} &=& \frac{d}{ds_j} \int dG^* G^* P(G^*|c) \\
	&=& \sum_m \frac{d}{ds_j} \pi_m \int dG^* G^* P(G^*|c, \bar{\bf X}^*_m) \\
\label{eq:dGdsm}
	&=& \sum_m \left( \pi_m \frac{d\bar{G}^*_m}{ds_j} + \frac{d\pi_m}{ds_j} \bar{G}^*_m \right).
\end{eqnarray}
The expressions for the individual $d\bar{G}^*_m/ds_j$ are given by
Eq.\ (\ref{eq:dGds24}), so the first term in Eqn.\ (\ref{eq:dGdsm})
exhibits direct functionality.  If the weights $\pi_m$ do not depend
appreciably on the $s_j$, the second term will be small, and the
restriction to direct functionality will be maintained.  If, on the
other hand, the weights do change appreciably (an obvious case might
be the presence of a bifurcation at a particular value of $s_j$), then
the second term may overpower the first enough to change the sign of
$\Delta \bar{G}^*/\Delta s$ and violate the restriction to direct
functionality.

We investigate this effect in two ways.  First, we show analytically
that, in the case of a 1-cycle, crossing a bifurcation does not
violate direct functionality.  Second, we subject all
positive-feedback networks to a numerical test to estimate the
dependence of the weights $\pi_m$ on the $s_j$.  The results of both
techniques suggest that the likelihood of a violation of direct
functionality due to the presence of multiple fixed points is low.

\subsection{Bifurcations do not violate direct functionality (1-D)}
\begin{figure}
\centering
\subfigure[]{\includegraphics [scale=.45] {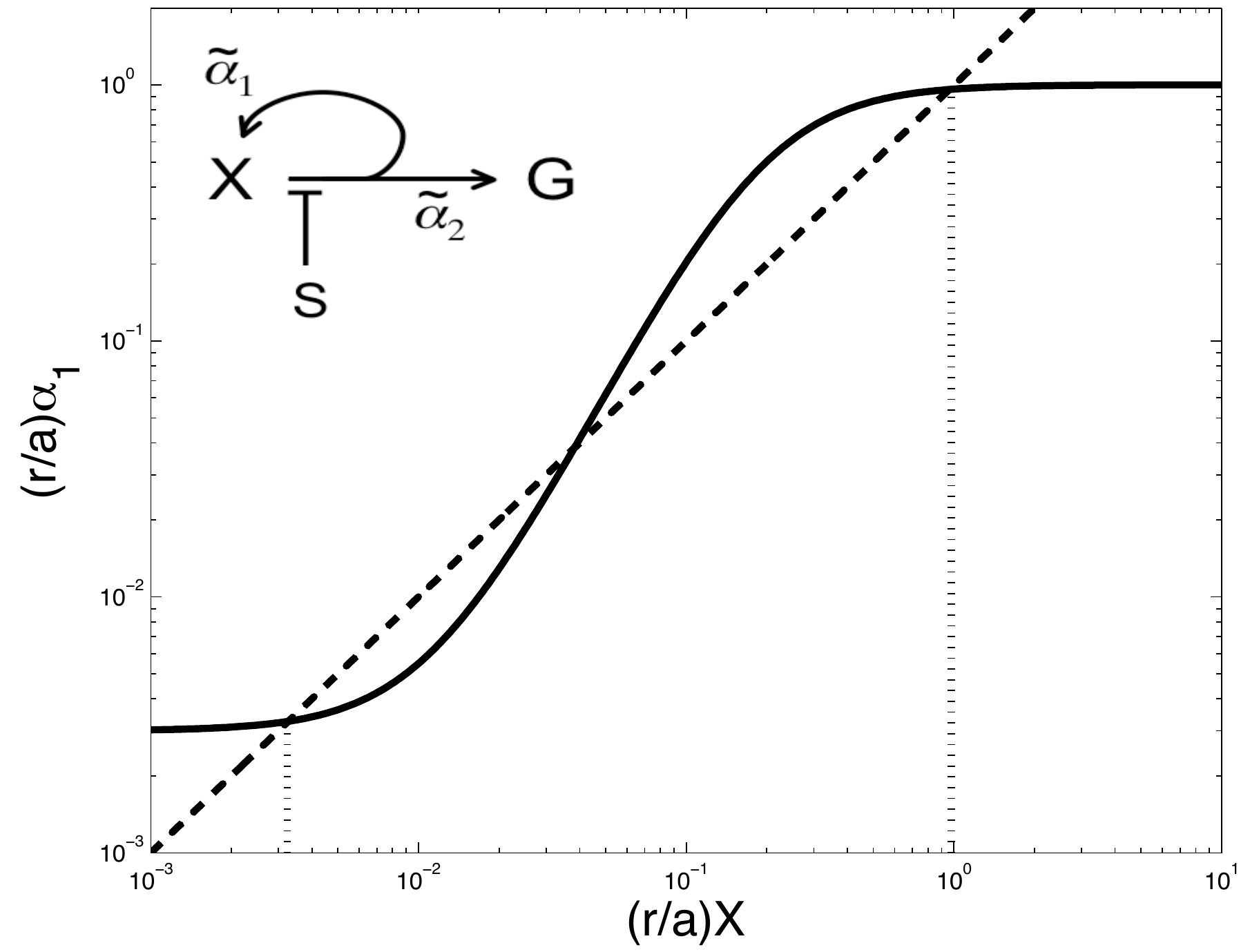} }
\subfigure[]{\includegraphics [scale=.45] {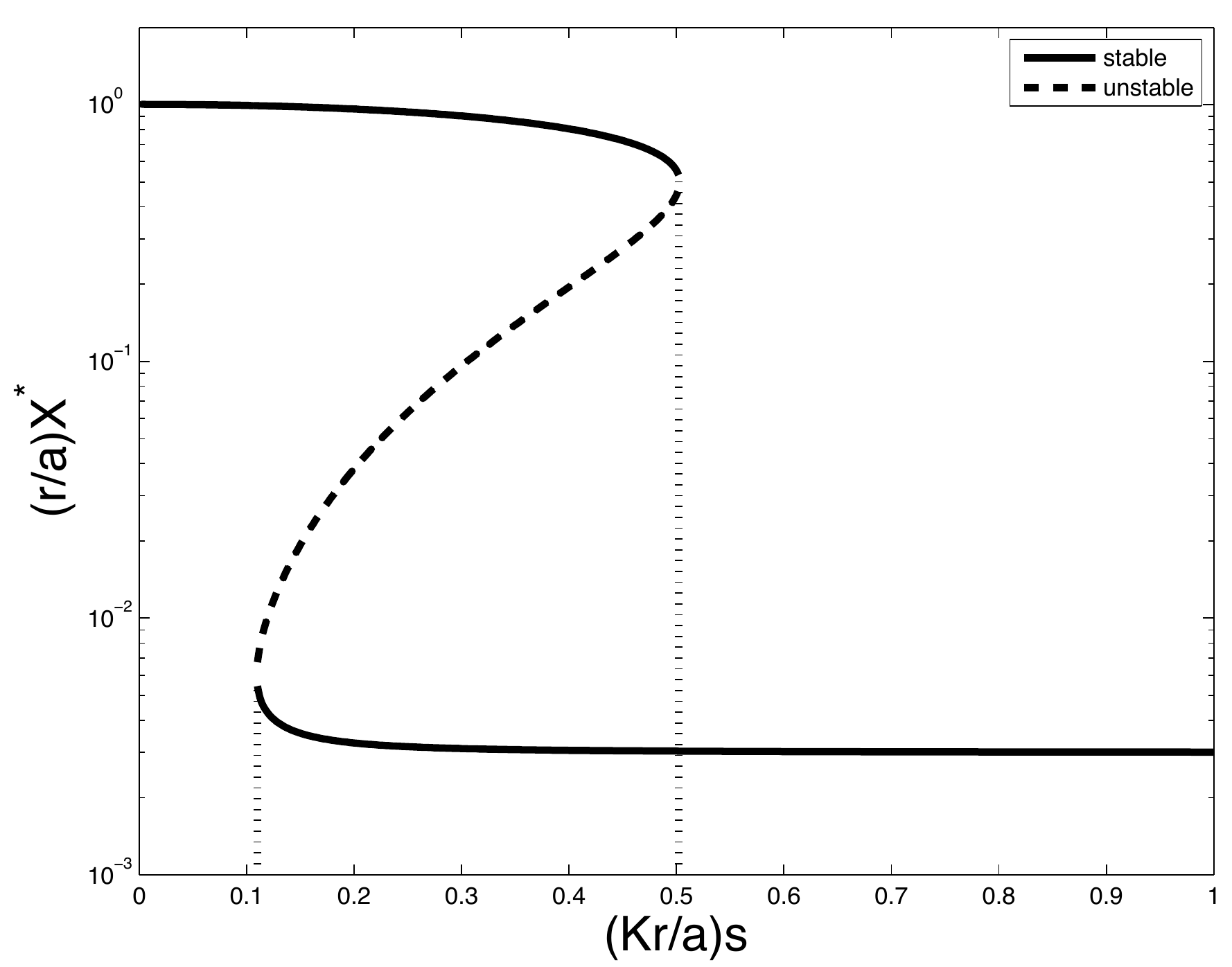} }
\caption{{\bf (a)} Solid: plot of regulation function $\alpha_1 =
  \tilde{\alpha}_1/r$ (refer to inset network), as defined in Eqns.\
  (\ref{eq:hill1}) and (\ref{eq:chi}), with parameters $h=2$,
  $a_0=0.03$, $a=10$, $K=2$, and $r=1$.  Dashed line shows $\alpha_1 =
  \bar{X}$ such that dotted lines indicate locations of stable fixed
  points $\bar{X}_1^*$ and $\bar{X}_2^*$ (take $\bar{X}_2^* >
  \bar{X}_1^*$).  {\bf (b)} Stable fixed points $\bar{X}_m^*$ (solid)
  and unstable fixed points (dashed) as a function of $s$, with $a_0/a
  = 0.003$ as in (a).  Dotted lines indicate locations of bifurcation
  points $s_1$ and $s_2$ such that only $\bar{X}_2^*$ exists when $s <
  s_1$, only $\bar{X}_1^*$ exists when $s > s_2$, and both
  $\bar{X}_1^*$ and $\bar{X}_2^*$ exist for $s_1 < s < s_2$.}
\label{fig:1cyc}
\end{figure}

Consider the case of a positive 1-cycle with a gene G immediately
outside, as shown in Fig.\ \ref{fig:1cyc}a (inset).  For $n=1$,
Eqns.\ (\ref{eq:fp1}-\ref{eq:fp3}) become
\begin{eqnarray}
\bar{X}^* &=& \alpha_1(\bar{\chi}^*),\\
\label{eq:fp2_1}
\bar{G}^* &=& \alpha_2(\bar{\chi}^*),
\end{eqnarray}
where unnecessary subscripts are dropped and $\bar{\chi} = \bar{X}/s$
as in Eqn.\ (\ref{eq:chi}).  With $\alpha_1$ of the form in Eqn.\
(\ref{eq:hill1}), there are at most two stable fixed points
$\bar{X}_1^*$ and $\bar{X}_2^*$, with $\bar{X}_1^* < \bar{X}_2^*$, as
illustrated by an example in Fig.\ \ref{fig:1cyc}a.  As shown in Fig.\
\ref{fig:1cyc}b, bifurcations occur at $s_1$ and $s_2$ such that only
$\bar{X}_2^*$ exists when $s < s_1$, only $\bar{X}_1^*$ exists when $s
> s_2$, and $\bar{X}_1^*$ and $\bar{X}_2^*$ are found with (unknown)
probabilities $\tilde{\pi}_1(s)$ and $\tilde{\pi}_2(s) = 1 -
\tilde{\pi}_1(s)$ respectively when $s_1 < s < s_2$.  These statements
can be combined such that
\begin{equation}
\pi_1(s) = \theta(s-s_1)\theta(s_2-s)\tilde{\pi}_1(s) + \theta(s-s_2)
\end{equation}
and $\pi_2(s) = 1 - \pi_1(s)$ define the probabilities of approaching
$\bar{X}_1^*$ and $\bar{X}_2^*$ respectively for any $s$.  Here $\theta$ is the Heaviside function.

As we go from an ``off'' value $s^-$ to an ``on'' value $s^+$, let
us assume that we hit both bifurcations, such that $s^- < s_1 < s_2 <
s^+$.  To test for direct functionality, we investigate the sign of
\begin{eqnarray}
  \frac{\Delta \bar{G}^*}{\Delta s} &=&
  \frac{1}{\Delta s}\int_{s^-}^{s^+} \sum_m \pi_m \frac{d\bar{G}^*_m}{ds} \, ds \nonumber \\
  && + \frac{1}{\Delta s}\int_{s^-}^{s^+} \sum_m \frac{d\pi_m}{ds} \bar{G}^*_m \, ds \\
  &\equiv& T_1 + T_2,
\end{eqnarray}
obtained using Eqns.\ (\ref{eq:deltaG}) and (\ref{eq:dGdsm}).  The
first term $T_1$ depends on
\begin{equation}
\frac{d\bar{G}^*_m}{ds} = \frac{\dot{\alpha_2}}{1-\alpha'_1}
\end{equation}
(from Eqn.\ (\ref{eq:dGds_1}); $\alpha' \equiv \partial
\alpha/\partial \bar{X}$ and $\dot{\alpha} \equiv \partial
\alpha/\partial s$ as before, both evaluated at the $m$th fixed
point), which, as previously discussed, is always of the sign of
$\dot{\alpha}_2$, consistent with direct functionality.

The second term $T_2$ can be written
\begin{eqnarray}
  T_2 &=& 	\frac{1}{\Delta s}\int_{s^-}^{s^+} \left( \frac{d\pi_1}{ds} \bar{G}^*_1 + 
    \frac{d\pi_2}{ds} \bar{G}^*_2 \right) \, ds \\
  &=& 	\frac{1}{\Delta s}\int_{s^-}^{s^+} -\frac{d\pi_1}{ds} (\bar{G}^*_2 - \bar{G}^*_1) \, ds,
\end{eqnarray}
and since
\begin{eqnarray}
\frac{d\pi_1}{ds} &=& \theta(s_2-s)\tilde{\pi}_1(s)\delta(s-s_1) \nonumber \\
	&& + \left[1 - \theta(s-s_1)\tilde{\pi}_1(s)\right]\delta(s-s_2) \nonumber \\
	&& + \theta(s-s_1)\theta(s_2-s)\frac{d\tilde{\pi}_1}{ds},
\end{eqnarray}
(where $\delta$ is the Dirac delta function) we have
\begin{eqnarray}
\label{eq:T2_1}
T_2 &=& \frac{1}{\Delta s} \left\{ -\left[ \tilde{\pi}_1(\bar{G}^*_2 - \bar{G}^*_1)\right]_{s_1} -
	\left[ \tilde{\pi}_2(\bar{G}^*_2 - \bar{G}^*_1)\right]_{s_2} \right. \nonumber \\
	&& \left. - \int_{s_1}^{s_2} \frac{d\tilde{\pi}_1}{ds} (\bar{G}^*_2 - \bar{G}^*_1) \, ds \right\}.
\end{eqnarray}
The first two terms in Eqn.\ (\ref{eq:T2_1}) represent the
contributions from crossing the bifurcations at $s_1$ and $s_2$
respectively.  Using Eqn.\ (\ref{eq:fp2_1}) we may write them as
\begin{eqnarray}
\label{eq:T2_2}
T_2 &=& \frac{1}{\Delta s}
\left\{\sum_{m=1}^2\left[ \tilde{\pi}_m\left(-\frac{\Delta\alpha_2}{\Delta\bar{\chi}^*}\right)
    \Delta\bar{\chi}^*\right]_{s_m} \right. \nonumber \\
&& \left. - \int_{s_1}^{s_2} \frac{d\tilde{\pi}_1}{ds} (\bar{G}^*_2 - \bar{G}^*_1) \, ds \right\},
\end{eqnarray}
where $\Delta \alpha_2 = \alpha_2(\bar{\chi}_2^*) -
\alpha_2(\bar{\chi}_1^*)$ and $\Delta \bar{\chi}^* = \bar{\chi}_2^* -
\bar{\chi}_1^* = (\bar{X}_2^* - \bar{X}_1^*)/s > 0$.  Since $\alpha_2$
is monotonic in $\bar{X}$, $-\Delta \alpha_2 / \Delta \bar{\chi}^*$ at
fixed $s$ is of the same sign as $-\alpha'_2$, which is of the same
sign as $\dot{\alpha}_2$ since $s$ effectively reduces $X$ (Eqn.\
(\ref{eq:chi})).  Therefore the contributions to $\Delta
\bar{G}^*/\Delta s$ from crossing the bifurcations do not violate
direct functionality.  A violation, at least in the case of a 1-cycle,
can only come from variations in the probabilities $\tilde{\pi}_m$
within the region $s_1 < s < s_2$, as described by the last term in
Eqn.\ (\ref{eq:T2_2}).  Next we describe a numerical test that
suggests such violations are rare.

\subsection{Numerics suggest violations from multiple fixed points are
  rare}
For each of the 12 positive-feedback networks, we numerically found
the steady state of the dynamical system with randomly sampled
parameters as before (cf.\ Numerical Results).  However now for each
parameter set we solved the system many times with randomly selected
initial conditions.  When multiple fixed points were found, the
fraction of trials approaching the $m$th fixed point was used for the
weight $\pi_m$.  This assumes the $\pi_m$ are determined only by the
basins of attraction of each fixed point, and by the distribution of
the initial conditions. However, different distributions of initial
conditions do not result in qualitative different results.

For each network, $2,000$ parameter sets were selected (uniform
randomly in log-space), at which the system was solved $100$ times
with initial protein counts selected uniform randomly from $0$ to
$1,000$ proteins per cell.  Over all positive-feedback networks,
$37\%$ of the parameter sets supported multiple fixed points for at
least one of the settings of S$_1$ and S$_2$.  However only $0.46\%$
of parameter sets produced violations of direct functionality.
Moreover this number is likely an overestimate, as no
distinguishability criterion was imposed as was done in the
single-fixed point case (cf.\ Numerical Results).  It is likely that
this fraction would remain low if the estimation of the $\pi_m$ was
refined to incorporate the curvatures of the fixed points, or if
alternative distributions were used for the sampling.

\section{All networks with serial regulation exhibit only direct
  functionality}
\label{sec:general}

In this section, we extend our analytic constraint as derived in the
context of the system studied experimentally by Guet et al.\
\cite{Guet:2002p38} to show that any network with only serial
regulation---each node having 0 or 1 parent---exhibits only direct
functionality, i.e.\, any target node X$_i$ changes with any input
S$_j$ according to the direct path between them.

We first consider a connected directed graph in which every node has
in-degree 1, called a {\it contrafunctional graph} \cite{harary}.  One
can show that a contrafunctional graph has exactly one cycle, each of
whose nodes is the root of a tree if the cycle edges are ignored
\cite{harary}.  Now consider changing one node's in-degree to 0, or
equivalently, removing an edge.  If the edge is in the cycle, the
graph remains connected and becomes a tree.  If the edge is not in the
cycle, the graph is cut into two components: a contrafunctional graph
and a tree.

A tree exhibits only direct functionality since there is at most one
path from an input S$_j$ to a gene X$_i$, which is therefore the
direct path.

In a contrafunctional graph, we first consider the case where the
target node X$_i$ is inside the cycle.  Only inputs S$_j$ that are
inside the cycle can affect X$_i$ because the rest of the graph
consists of trees that all point away from the cycle.  Since we can
start labeling nodes at any point in the cycle, we may take $i \leq j$
without loss of generality.  Then, using the chain rule,
\begin{eqnarray}
\frac{d\bar{X}_i^*}{ds_j} &=&
	\left[ \dot{\alpha}_{(j \, {\rm mod} \, n)+1}
	+ \alpha'_{(j \, {\rm mod} \, n)+1} \frac{d\bar{X}_j^*}{ds_j} \right] \nonumber \\
	&& \times \frac{\prod_{k=1}^n \alpha'_k}{\prod_{l=i}^j \alpha'_{(l \, {\rm mod} \, n)+1}}\\
\label{eq:dXidsj}
&=& \left[ \frac{1}{1 - \prod_{m=1}^n \alpha'_m} \right] \nonumber \\
	&& \times \dot{\alpha}_{(j \, {\rm mod} \, n)+1}
	\frac{\prod_{k=1}^n \alpha'_k}{\prod_{l=i}^j \alpha'_{(l \, {\rm mod} \, n)+1}},
\end{eqnarray}
where the second step follows from Eqn.\ (\ref{eq:dXjdsj}).

We next consider the case where the target node is outside the cycle.
An input S$_j$ can only affect the node if it is either in the cycle
or above the node in its tree.  The portion of the path in the tree
will exhibit direct functionality.  Therefore in looking for possible
indirect functionality we may, without loss of generality, take the
node to be immediately outside the cycle, as we did for G in the
previous section.  $d\bar{G}^*/ds_j$ is then given by Eqn.\
(\ref{eq:dGds_1}).

In both Eqns.\ (\ref{eq:dXidsj}) and (\ref{eq:dGds_1}), the term
outside the brackets represents the direct path from S$_j$ to the
target node, and the term inside the brackets is positive for stable
fixed points.  Therefore, a contrafunctional graph exhibits only
direct functionality.  Since each connected component of a network in
which every node has in-degree 0 or 1 is either a contrafunctional
graph or a tree, such networks exhibit only direct
functionality. Thus, in general, the possible logical functions of
topologies with at most one regulator per node are severely
constrained.

\section{Appendix: The stochastic model}
\label{sec:model}

The dynamical system in Eqns.\ (\ref{eq:ode1}-\ref{eq:odeG}) provides
a deterministic description of mean expression levels but fails to
capture fluctuations around these means.  A full stochastic
description is given by the chemical master equation.  For $N$ species
participating in $R$ elementary reactions in a system with volume
$\Omega$, the master equation reads
\begin{equation}
\label{eq:master}
\frac{dP({\bf n}, t)}{dt} = \Omega \sum_{j=1}^R \left( \prod_{i = 1}^N E^{-Z_{ij}} - 1\right)
	f_j({\bf X}, \Omega) P({\bf n}, t),
\end{equation}
where $P({\bf n}, t)$ is the probability of having the copy number
vector ${\bf n} = \Omega {\bf X} = \Omega (X_1, \dots, X_N )$ at time
$t$, $Z_{ij}$ is the $N \times R$ stochiometric matrix,
$E^{-Z_{ij}}$ is the step operator which acts by removing
$Z_{ij}$ molecules from $n_i$, and $f_j$ is the rate for reaction
$j$.  The $f_j$ are the $\tilde{\alpha}_j$ and $r_j X_j$ of Eqns.\ (\ref{eq:ode1}-\ref{eq:odeG}) in the
macroscopic limit.

As in previous work \cite{Ziv:2007p88}, we employ the much-used {\it
  linear noise approximation} \cite{Elf:2003p86, Paulsson:2004p85,
  Elf:2003p87, vanKampen} to make Eqn.\ (\ref{eq:master}) tractable by
expanding in orders of $\Omega^{-1/2}$.  Introducing ${\bf \xi}$ such
that $n_i = \Omega X_i + \Omega^{1/2} \xi_i$ and treating ${\bf \xi}$
as continuous, the first two terms in the expansion yield the
macroscopic rate equations (e.g. Eqns.\ (\ref{eq:ode1}-\ref{eq:odeG})
in our case) and the linear Fokker-Plank equation, respectively:
\begin{eqnarray}
\sum_{i=1}^N \frac{\partial \bar{X}_i}{\partial t} \frac{\partial P({\bf \xi}, t)}{\partial \xi_i}
	= \sum_{i=1}^N \sum_{j=1}^R Z_{ij} f_j({\bf \bar{X}}) \frac{\partial P({\bf \xi}, t)}{\partial \xi_i}, \\
\label{eq:fokker}
\frac{\partial P({\bf \xi}, t)}{\partial t} =
	- \sum_{i, k} J_{ik} \frac{\partial (\xi_k P)}{\partial \xi_i}
	+ \frac{1}{2} \sum_{i, k} D_{ik} \frac{\partial^2 P}{\partial \xi_i \partial \xi_k},
\end{eqnarray}
where $J_{ik} = \sum_{j=1}^R Z_{ij} (\partial f_j/\partial X_k)$
is the Jacobian matrix (e.g. Eqn.\ (\ref{eq:jacobian})) and $D_{ik} =
\sum_{j=1}^R Z_{ij} Z_{kj} f_j({\bf X})$ is a diffusion-like
matrix.  The steady-state solution to Eqn.\ (\ref{eq:fokker}) is the
multivariate Gaussian
\begin{equation}
P({\bf \xi}) = \left[ (2\pi)^N \det \Xi \right]^{-1/2} \exp \left( - \frac{-\xi^T \Xi \xi}{2} \right),
\end{equation}
where the covariance matrix $\Xi$ satisfies
\begin{equation}
\label{eq:lyap}
J \Xi + \Xi J^T + D = 0.
\end{equation}
We solve for $\Xi$ using standard matrix Lyapunov equation solvers
(e.g., Matlab's \texttt{lyap}).  Thus fluctuations are captured to
leading order by Gaussian distributions with means $\bar{X}_i$ given
by the macroscopic equation and variances given by the diagonal
entries of $\Xi$.  For example, Gaussian distributions $P(G^*|c)$ are
shown in Fig.\ \ref{fig:func}b for the steady-state concentration of
the reporter gene G under chemical input states $c$. In
\cite{Ziv:2007p88} we have compared the distributions $P(G^*|c)$
obtained using the linear noise approximation to those obtained via
direct stochastic simulations \cite{Gillespie:1977p54} and found the
results almost indistinguishable for molecular copy number above
10-20.

\acknowledgements 

We are grateful to the organizers and participants of The First q-bio
Conference, where a preliminary version of this work was presented. This
work was partially supported by NSF Grant No.\ ECS-0425850 to CW and
IN. IN was further supported by LANL LDRD program under DOE Contract
No.\ DE-AC52-06NA25396.

\bibliographystyle{apsrev}
\bibliography{qbio_paper4}

\end{document}